\documentclass[10pt,journal]{IEEEtran}

\usepackage[final]{graphicx}
\usepackage{epsfig}
\usepackage{caption2}
\usepackage{amsmath}
\usepackage{amssymb,times}
\usepackage{bm}
\usepackage{color}
\usepackage{cite}
\usepackage[ruled,vlined]{algorithm2e}     
\usepackage{epstopdf}

\newcommand{\ba}{\begin{array}}
\newcommand{\ea}{\end{array}}
\newcommand{\be}{\begin{displaymath}}
\newcommand{\ee}{\end{displaymath}}
\newcommand{\ben}{\begin{equation}}
\newcommand{\een}{\end{equation}}
\newcommand{\bena}{\begin{eqnarray}}
\newcommand{\eena}{\end{eqnarray}}
\newcommand{\beqa}{\begin{eqnarray*}}
\newcommand{\enqa}{\end{eqnarray*}}

\newcommand{\bc}{\begin{center}}
\newcommand{\ec}{\end{center}}
\newcommand{\bi}{\begin{itemize}}
\newcommand{\ei}{\end{itemize}}
\newcommand{\benu}{\begin{enumerate}}
\newcommand{\eenu}{\end{enumerate}}
\newcommand{\bdes}{\begin{description}}
\newcommand{\edes}{\end{description}}
\newcommand{\bt}{\begin{tabular}}
\newcommand{\et}{\end{tabular}}

\newcommand \thetabf{{\mbox{\boldmath$\theta$\unboldmath}}}

\newcommand \alphabf{\mbox{\boldmath$\alpha$\unboldmath}}

\newcommand \Cbf{{\bf C}}

\newcommand{\circlambda}{\mbox{$\Lambda$
             \kern-.85em\raise1.5ex
             \hbox{$\scriptstyle{\circ}$}}\,}

\renewcommand \thetabf{\boldsymbol{\theta}}

\renewcommand \alphabf{\boldsymbol{\alpha}}

\IEEEoverridecommandlockouts

\begin{document}

\title{Statistical Performance of Generalized Direction Detectors with Known Spatial Steering Vector}
\author{Zhenyu Xu, Weijian Liu, \emph{Senior Member, IEEE}, Changfei Wu, Ming Liu\\ Qinglei Du, and Jun Liu, \emph{Senior Member, IEEE} 

\thanks{This manuscript is accepted by IEEE Signal Processing Letters. First received February 15, 2025; revised April 22, 2025; accepted May 3, 2025. 
This work was supported in part by the National Natural Science Foundation of China under Contracts 62071482,
62471485, 62471450, the Natural Science Foundation of Hubei Province under Grant 2025AFB873, and the Natural Science Foundation of Anhui Province under Grant 2208085J17.
The associate editor coordinating the review of this article and approving it for publication was Dr Wenqiang Pu. \emph{(Corresponding author: Weijian Liu.)}
}
\thanks{Z. Xu, W. Liu, C. Wu, M. Liu, and Q. Du are with Wuhan Electronic Information Institute, Wuhan 430019, China (e-mails: 1935620989@qq.com, liuvjian@163.com, wuchangfei21@163.com, liuminglove1001@163.com, and dql822@163.com).}
\thanks{J. Liu is with Department of Electronic Engineering and Information Science, University of Science and Technology of China, Hefei,  {\rm 230027}, China (e-mail: junliu@ustc.edu.cn).}

}

\maketitle

\begin{abstract}
The generalized direction detection (GDD) problem involves determining the presence of a signal of interest within matrix-valued data, where the row and column spaces of the signal (if present) are known, but the specific coordinates are unknown. Many detectors have been proposed for GDD, yet there is a lack of analytical results regarding their statistical detection performance. This paper presents a theoretical analysis of two adaptive detectors for GDD in scenarios with known spatial steering vectors. Specifically, we establish their statistical distributions and develop closed-form expressions for both detection probability (PD) and false alarm probability (PFA). Simulation experiments are carried out to validate the theoretical results, demonstrating good agreement between theoretical and simulated results.

\end{abstract}


\section{Introduction}
\IEEEPARstart{M}{ULTICHANNEL} signal detection in unknown noise is one of the fundamental issues in modern signal processing \cite{LiuLiu22SCIS}. In recent years, numerous multichannel adaptive detection problems have been studied and many adaptive detectors have been proposed \cite{ChenGai22,Besson23SPL,GuanMu23,ZhuZhu23,LiuLiu23SPL, XueFan24,MiaoZhang24,ShenLiu25SPL,LiuWu25SCIS}. Among these problems, directional detection (DD) is one of the categories. Its aim is to detect whether there exists a target within matrix-valued data. The signal components backscattered from the potential target all originate from the identical direction. However, the precise direction is unknown, and it is only known that the signal components are in a certain subspace.
Consequently, the signal steering vector resides within a predefined subspace of more than one dimension, but its specific coordinates within this space are unknown. To tackle the DD problem in partially homogeneous scenarios, the authors in \cite{BessonScharf06b} developed the generalized adaptive direction detector (GADD), which operates under the condition that the noise covariance matrices of the test data and the training data vary solely by an unascertained scaling coefficient. The DD problem was also explored in \cite{BandieraBesson07}, with the assumption that there exists a certain subspace interference. Then, utilizing the generalized likelihood ratio test (GLRT) and its two-step adaptation, two direction detection methods were developed for homogeneous settings, assuming identical noise covariance matrices for both test and training data. The corresponding Wald test was given in \cite{LiLiu18MSSP}. The model in \cite{BandieraBesson07} was generalized in \cite{BandieraBesson13} and \cite{DongLiu17SPL_DD_PHE}. Specifically, in \cite{BandieraBesson13}, the interference subspace is not known, and the detection problem was addressed using an ad hoc algorithm that has a close connection to the GLRT. In \cite{DongLiu17SPL_DD_PHE}, the environment is partial homogeneity, and the problem was resolved according to the two-step GLRT and two-step Wald test.

The problem of DD can be further extended to the case of generalized direction detection (GDD). In this situation, it is assumed that the row space and column space of a rank-one matrix-valued signal are known, but the specific coordinates within these spaces are not. In \cite{BoseSteinhardt96a}, the GLRT was derived for the GDD when there is no training data and the dimensions of the signal matrices are restricted to meet specific conditions. When dealing with data of limited dimensionality, utilizing a unitary matrix transformation on the test data enables the creation of a corresponding virtual training dataset that is functionally equivalent. Nonetheless, this process entails the sacrifice of the signal's row structure. Mathematically, the data model presented in \cite{BoseSteinhardt96a} exhibits equivalence to the one described in \cite{LiuXie14c}, where a direction detector was provided according to the Wald test.

In \cite{LiuLiu2015b}, the problem of GDD was addressed in the context of having training data that share the same noise covariance matrix with the test data. Based on the GLRT, two detectors were developed in \cite{LiuLiu2015b}. The corresponding Wald test for the detection problem in \cite{LiuLiu2015b} was obtained from \cite{Liu20SCIS}. The results in \cite{LiuLiu2015b} were further expanded in \cite{LiuZhang21} in the case of limited training data.
In \cite{LiuZhang21}, after performing a unitary transformation on the test data, the reduced-dimension test data and virtual training data were obtained. The virtual training data, together with the real training data, were jointly used for the estimation of the unknown noise covariance matrix, thus reducing the requirement for the training data. Moreover, the GDD problem was also considered in \cite{Liu19SCIS} for the partially homogeneous environment.

Although the GDD problem has been investigated in the above references, the statistical performance of the corresponding detectors remains unexplored in the existing literature. In this letter, we determine the statistical distributions of the GLRT-based detectors assuming the spatial steering vector is known. This scenario arises when there is no uncertainty on the spatial information about the target. Such a simplification makes theoretical analysis more tractable and lays the foundation for studying more complex scenarios. For this case, we readily obtain the expressions for the probabilities of detection (PDs) and probabilities of false alarm (PFAs), which are verified through simulation experiments.

The structure of this letter is outlined below. Section II discusses the detection problem and introduces the associated detectors. Section III derives the statistical performance of the detectors as well as the computational expressions for the PDs and PFAs. Section IV verifies the above theoretical results based on simulation experiments. Finally, Section V provides a summary of the contributions presented in this letter.


\section{Problem Formulation}

An $O\times P$ test data $\mathbf{Z}$ solely comprises noise $\mathbf{V}$ under hypothesis $\text{H}_0$. In contrast,  $\mathbf{Z}$ comprises noise $\mathbf{V}$ and signal $\mathbf{H}$ under hypothesis $\text{H}_1$. The possible signal $\mathbf{H}$ has the form $\mathbf{H}=\gamma\mathbf{a}\mathbf{b}^H$, where $\gamma$ is the signal amplitude, $\mathbf{a}$ is the spatial steering vector, $\mathbf{b}$ is the temporal steering vector, and $(\cdot)^H$ is the conjugate transpose.
When there exist uncertainties in $\mathbf{a}$ and $\mathbf{b}$, it is assumed that $\mathbf{a}$ and $\mathbf{b}$ reside within known subspaces while having unknown coordinates. Hence, we can modify the signal model as \begin{equation}
\label{}
\mathbf{H}={\mathbf{A}}\bm\theta {\bm\alpha ^H}{\mathbf{C}},
\end{equation}
where the $O\times J$ matrix $\mathbf{A}$ and $Q\times P$ matrix $\mathbf{C}$ are known signal matrices, while $\bm\theta$ and $\bm\alpha$ are unknown coordinates.
The GDD problem can be formulated as \cite{LiuLiu2015b,LiuZhang21,Liu19SCIS}:
\begin{equation}
\label{hypothesis}
\left\{ \begin{array}{l}
{\text{H}_0}:{\mathbf{Z}} = {\mathbf{V}}, \quad {{\mathbf{Z}}_L} = {{\mathbf{V}}_L},\\
{\text{H}_1}:{\mathbf{Z}} = {\mathbf{A}}\bm\theta {\bm\alpha ^H}{\mathbf{C}} + {\mathbf{V}},\quad {{\mathbf{Z}}_L} = {{\mathbf{V}}_L}.
\end{array} \right.
\end{equation}
where the $O\times P$ matrix $\mathbf{Z}$ is the test data, the $O\times L$ matrix $\mathbf{Z}_L$ is the training data, $\mathbf{V}$ and $\mathbf{V}_L$ are the noise in $\mathbf{Z}$ and $\mathbf{Z}_L$, respectively.
The columns of $\mathbf{A}$ and the rows of $\mathbf{C}$ span the column subspace and row subspace, respectively. $\bm\theta$ and $\bm\alpha$ denote the corresponding coordinates. 
In addition to the context of GDD, the detection model in \eqref{hypothesis} is also addressed in other contexts, such as \cite{JianLiu22DSP}.

The GLRT and 2S-GLRT for the detection problem in \eqref{hypothesis} are given in \cite{LiuLiu2015b,LiuZhang21}. To the best of our knowledge, no reference derived their statistical performance. In this letter, we derive the statistical performance in the case of known spatial steering vector, i.e., $J=1$, where the matrix $\mathbf{A}$ in \eqref{hypothesis} becomes a column vector, denoted as $\mathbf{a}$.
In this case, the GLRT and 2S-GLRT are
\begin{equation}
\label{30}
\begin{array}{r}
\begin{aligned}
{t_\text{GLRGDD}} = &{{{\left[ {{{\mathbf{a}}^H}{{({\mathbf{S}} + {\mathbf{Z}}{{\mathbf{Z}}^H})}^{ - 1}}{\mathbf{a}}} \right]}^{ - 1}}} \\
 &\cdot {{\mathbf{a}}^H}{{\mathbf{S}}^{ - 1}}{\mathbf{Z}}{({{\mathbf{I}}_P} + {{\mathbf{Z}}^H}{{\mathbf{S}}^{ - 1}}{\mathbf{Z}})^{ - 1}}{{\mathbf{C}}^H}\\
 &\cdot {\left[ {{\mathbf{C}}{{({{\mathbf{I}}_P} + {{\mathbf{Z}}^H}{{\mathbf{S}}^{ - 1}}{\mathbf{Z}})}^{ - 1}}{{\mathbf{C}}^H}} \right]^{ - 1}}\\
 &\cdot {\mathbf{C}}{{{({{\mathbf{I}}_P} + {{\mathbf{Z}}^H}{{\mathbf{S}}^{ - 1}}{\mathbf{Z}})}^{ - 1}}{{\mathbf{Z}}^H}{{\mathbf{S}}^{ - 1}}{\mathbf{a}}} .
\end{aligned}
\end{array}
\end{equation}
and
\begin{equation}
\label{23}
{t_\text{AMGDD}} =  \frac{{\mathbf{a}}^H{{\mathbf{S}}^{ - 1}}{\mathbf{Z}} {{\mathbf{P}}_{{{\mathbf{C}}^H}}}{{\mathbf{Z}}^H}{{\mathbf{S}}^{ - 1}}{\mathbf{a}}} {{{\mathbf{a}}^H}{{\mathbf{S}}^{ - 1}}{\mathbf{a}}}
\end{equation}
respectively, where ${{\mathbf{P}}_{{{\mathbf{C}}^H}}}={\mathbf{C}}^H ({\mathbf{C}}{\mathbf{C}}^H) ^{-1}{\mathbf{C}}$ and ${\bf{S}}={\bf{Z}}_L{\bf{Z}}_L^H$. 
The above two detectors in \eqref{30} and \eqref{23} are referred to as the generalized likelihood ratio-based generalized direction detector (GLRGDD) and adaptive matched  generalized direction detector (AMGDD), respectively in \cite{LiuLiu2015b}.

\section{Statistical performance}
Using the results in \cite{LiuZhang21}, we can rewrite \eqref{30} as
\begin{equation}
\label{12}
\begin{array}{c}
{t_\text{GLRGDD}} = \dfrac{{{{\mathbf{a}}^H}{\mathbf{S}}_ + ^{ - 1}{{\mathbf{Z}}_*}} {{{({{\mathbf{I}}_Q} + {\mathbf{Z}}_*^H{\mathbf{S}}_ + ^{ - 1}{{\mathbf{Z}}_*})}^{ - 1}}}
{{\mathbf{Z}}_*^H{\mathbf{S}}_ + ^{ - 1} {\mathbf{a}} }}
{{{{\mathbf{a}}^H}{\mathbf{S}}_ + ^{ - 1}{\mathbf{a}}}}
\end{array}
\end{equation}
where $\mathbf{S}_+=\mathbf{S}+\mathbf{Z}\mathbf{P}_{\mathbf{C}^H}^\bot  \mathbf{Z}^H$, $\mathbf{P}_{\mathbf{C}^H}^\bot =\mathbf{I}_P- \mathbf{P}_{\mathbf{C}^H} $, and $\mathbf{Z}_*= \mathbf{Z} \mathbf{C}^H{(\Cbf\Cbf_{}^H)^{-1/2}}$.
Equation \eqref{12} is equivalent to
\begin{equation}
\label{GLRGDDRU1}
\begin{array}{l}
t_\text{GLRGDD}^\prime =\\
  {\dfrac{{\mathbf{a}}^H{\mathbf{S}}_+^{-1} {{\mathbf{Z}}_*} {{{({{\mathbf{I}}_Q} + {\mathbf{Z}}_*^H{\mathbf{S}}_ + ^{ - 1}{{\mathbf{Z}}_*})}^{ - 1}}} {\mathbf{Z}}_*^H {\mathbf{S}}_ + ^{ - 1}{\mathbf{a}}}
{{{{\mathbf{a}}^H}{\mathbf{S}}_ + ^{ - 1}{\mathbf{a}}} - {{\mathbf{a}}^H{\mathbf{S}}_+^{-1} {{\mathbf{Z}}_*} {{{({{\mathbf{I}}_Q} + {\mathbf{Z}}_*^H{\mathbf{S}}_ + ^{ - 1}{{\mathbf{Z}}_*})}^{ - 1}}} {\mathbf{Z}}_*^H {\mathbf{S}}_ + ^{ - 1}{\mathbf{a}}}}},
\end{array}
\end{equation}
due to the fact that $t_\text{GLRGDD}^\prime=1/(t_\text{GLRGDD}^{-1}-1)$ can be regarded as a consistently rising function of $t_\text{GLRGDD}$. 

Equation \eqref{23} can be similarly expressed as
\begin{equation}
\label{AMGDD}
{t_\text{AMGDD}} = \frac{{{\mathbf{a}}^H}{\mathbf{S}} ^{ - 1}{{\mathbf{Z}}_*} {\mathbf{Z}}_*^H{\mathbf{S}} ^{ - 1}{\mathbf{a}}} {{{\mathbf{a}}^H}{\mathbf{S}} ^{ - 1}{\mathbf{a}}}
\end{equation}

Remarkably, the GLRGDD in \eqref{GLRGDDRU1} and the AMGDD in \eqref{AMGDD} share the same forms as the multi-band GLR (MBGLR) in \cite{WangCai91} 
and the generalized adaptive matched filter (GAMF) in \cite{ConteDeMaio01}, respectively. The statistical characteristics of the MBGLR were detailed in \cite{WangCai91,Raghavan13b}, whereas those of the GAMF were discussed in \cite{Raghavan13a}. By leveraging the findings from \cite{WangCai91,Raghavan13b,Raghavan13a}, we can easily derive the statistical properties of the GLRGDD in \eqref{GLRGDDRU1} and AMGDD in \eqref{AMGDD}.
Precisely, ${\mathbf{a}}$, ${\mathbf{S}}_{{+}}$ and ${{\mathbf{Z}}_*}$ in \eqref{GLRGDDRU1} and \eqref{AMGDD} serve as the signal steering vector, sample covariance matrix (SCM), and test data matrix in the detection statistic of the MBGLR and GAMF, respectively.
Hence, in a manner similar to \cite{WangCai91,Raghavan13b}, we have that the conditional statistical distribution of the GLRGDD in \eqref{GLRGDDRU1} under hypothesis $\text{H}_1$ is a complex non-central F-distribution with $P$ and $L-O+1$ degrees of freedom (DOFs) and a noncentrality parameter $\beta_{\text{G}}$, written symbolically as
\begin{equation}
\label{ch04_GLRT_HE_p1_cSD}
t_{\text{GLRGDD}}|\text{H}_1\sim{\cal C}{\cal F}_{Q,L+P-Q-O+1}\left(\rho\beta_{\text{G}}\right),
\end{equation}
where $\rho$ can be defined as the signal-to-noise ratio (SNR), given as
\begin{equation}
\label{ch04_SNR_p1}
\rho=|\theta|^2\alphabf^H{\mathbf{C}}{\mathbf{C}}^H\alphabf\cdot{\mathbf{a}}^H{\mathbf{R}}^{-1}{\mathbf{a}},
\end{equation}
with $\theta$ being a scalar version of the vector $\thetabf$ in \eqref{hypothesis}.
In \eqref{ch04_GLRT_HE_p1_cSD}, $\beta_{\text{G}}$ is random variable, which 
is distributed as a complex central Beta distribution, with $L+P-O+1$ and $O-1$ DOFs, written symbolically as 
\begin{equation}
\label{ch04_CB00}
\beta_{\text{G}}\sim {\cal C}{\cal B}_{L+P-O+1,O-1}. 
\end{equation}
Under hypothesis $\text{H}_0$, \eqref{ch04_GLRT_HE_p1_cSD} reduces to a complex central F-distribution, i.e., 
\begin{equation}
\label{ch04_GLRT_HE_p0_cSD}
t_{\text{GLRGDD}}|\text{H}_0\sim{\cal C}{\cal F}_{Q,L+P-Q-O+1}.
\end{equation}

The PD of the GLRGDD in \eqref{GLRGDDRU1} can be expressed as
\begin{equation}
\label{ch04_PD_GLRGDD}
\begin{array}{c}
\begin{aligned}
\text{PD}{_{\text{GLRGDD}}} &= \text{Pr}[t_{\text{GLRGDD}} > {\eta _{\text{G}}};{\text{H}_1}]\\
&= \int_0^1 [1 -{{\cal P}_{1}{({\eta_{\text{G}}})}}]{f}({\beta_{\text{G}}}) {\kern 1pt} d{\beta_{\text{G}}},
\end{aligned}
\end{array}
\end{equation}
where ${f}({\beta_{\text{G}}})$ is the probability density function (PDF) of the random variable ${\beta_{\text{G}}}$ in \eqref{ch04_CB00}, 
${\eta _{\text{G}}}$ is the detection threshold of the GLRGDD in \eqref{GLRGDDRU1},
${{\cal P}_{1}}(\eta_{\text{G}})$ is cumulative distribution function (CDF) of $t_\text{GLRGDD}$ under hypothesis $\text{H}_1$ for given $\beta_{\text{G}}$, namely, 
\begin{equation}
\label{ch04_CDF_GLRGDD--RU_1}
{{\cal P}_{1}}({\eta _{\text{GLRGDD}}}) = \text{Pr}\ [t_{\text{GLRGDD}} \le {\eta _{\text{G}}}|{\beta_{\text{G}}}; \text{H}_1].
\end{equation}
According to (A.29) in \cite{KellyForsythe89}, we have
\begin{equation}
\label{ch04_CDF_GLRGDD--RU1}
{{\cal P}_{ 1}}({\eta_{\text{G}}}) = \sum\limits_{k = 0}^{L+P-Q-O} {\text{C}_{L+P-O}^{k+Q}} \frac{\eta_{\text{G}}^{k+Q} {{\text{IG}_{k + 1}}\left(\frac{\rho\beta_{\text{G}}} {1 + \eta_{\text{G}} }\right)} } {{(1 + \eta_{\text{G}})}^{L+P-O}},
\end{equation}
where $\text{C}_n^m = \frac{n!}  {m!(n - m)!}$ is the binominal coefficient and $\text{IG}{_{k + 1}}(a) = {e^{ - a}}\sum\nolimits_{m = 0}^k {\frac{{a^m}} {m!}}$ is the incomplete Gamma function.
According to  (A2-12) in \cite{KellyForsythe89}, the PDF of $\beta_{\text{G}}$ is
\begin{equation}
\label{PDF}
f(\beta_{\text{G}})= \frac{{{\beta_{\text{G}} ^{L+P-O}}{{(1 - \beta_{\text{G}} )}^{O-2}}}}{\text{B}({{L+P-O+1},O-1})},
\end{equation}
where ${\text{B}({m,n})}=\frac{(m-1)!(n-1)!}{(m+n-1)!}$ is the Beta function with integer argument.
Substituting \eqref{ch04_CDF_GLRGDD--RU1} and \eqref{PDF} into \eqref{ch04_PD_GLRGDD}, we can derive the final expression for the PD of the GLRGDD in \eqref{GLRGDDRU1}, and setting $\rho=0$ in these results leads to the corresponding expression for the PFA.

Moreover, in a manner similar to \cite{Raghavan13a}, we can express the AMGDD in \eqref{AMGDD} as the ratio of two random variables, namely,
\begin{equation}
\label{PDAMGDDRU}
{t_{\text{AMGDD}}}= \frac{\kappa}{\beta_\text{A}}
\end{equation}
where the conditional statistical distribution of $\kappa$ under  $\text{H}_1$, with ${\beta_\text{A}}$ fixed, is
\begin{equation}
\label{ch04_AMGDD--RU_con_SD}
\kappa|\text{H}_1\sim{\cal CF}_{Q,L+P-Q-O+1}(\beta_{\text{A}}\rho),
\end{equation}
 and ${\beta_{\text{A}}}$ is distributed as
\begin{equation}
\label{ch04_beta_AMGDD--RU_SD}
\beta_{\text{A}}\sim{\cal CB}_{L+P-Q-O+2,O-1}. 
\end{equation}
Under  $\text{H}_0$, \eqref{ch04_AMGDD--RU_con_SD} reduces to 
\begin{equation}
\label{ch04_AMGDD--RU_con_SD0}
\kappa|\text{H}_0\sim{\cal CF}_{Q,L+P-Q-O+1}.
\end{equation}

The PD of the AMGDD in \eqref{AMGDD} can be expressed as
\begin{equation}
\label{PD:AMGDD--RU}
\begin{array}{c}
\begin{aligned}
\text{PD}{_{\text{AMGDD}}} &= \text{Pr}[t_{\text{AMGDD}} > {\eta _{\text{A}}};{\text{H}_1}]\\
&= \int_0^1 [1 -{{\cal P}_{1}{({\eta_{\text{A}}}\beta_{\text{A}})}}]{f} ({\beta_{\text{A}}}) {\kern 1pt} d{\beta_{\text{A}}},
\end{aligned}
\end{array}
\end{equation}
where $\eta_\text{A}$ is the detection threshold for the AMGDD, ${{\cal P}_{1}{({\eta_{\text{A}}}\beta_{\text{A}})}}$ is the CDF of $\kappa$ in \eqref{ch04_AMGDD--RU_con_SD}, ${f} ({\beta_{\text{A}}})$ is the PDF of ${\beta_{\text{A}}}$ in \eqref{ch04_beta_AMGDD--RU_SD}. Note that ${{\cal P}_{1}{({\eta_{\text{A}}}\beta_{\text{A}})}}$ can be obtained by replacing ${\eta_{\text{G}}} $ and $\beta_\text{G}$ with ${\eta_{\text{A}}}\beta_{\text{A}}$ and $\beta_{\text{A}}$ in \eqref{ch04_CDF_GLRGDD--RU1}, respectively, while ${f} ({\beta_{\text{A}}})$ can be similarly obtained from \eqref{PDF}, i.e.,
\begin{equation}
\label{PDFBetaA}
f(\beta_{\text{A}})= \frac{{{\beta_{\text{G}} ^{L+P-Q-O+1}}{{(1 - \beta_{\text{G}} )}^{O-2}}}} {\text{B}({{L+P-Q-O+2},O-1})}.
\end{equation}

Subsequently, by substituting \eqref{ch04_CDF_GLRGDD--RU1} and \eqref{PDFBetaA} into \eqref{PD:AMGDD--RU}, and replacing ${\eta_{\text{G}}} $ and $\beta_\text{G}$ with ${\eta_{\text{A}}}\beta_{\text{A}}$ and $\beta_{\text{A}}$ respectively, we obtain the PD of the AMGDD in \eqref{23}. Moreover, when $\rho = 0$ is set in these results, the expression for the corresponding PFA is obtained.


\section{Numerical Examples}
In this section, Monte Carlo simulations are carried out to corroborate the theoretically calculated expressions for the detection probability and the false alarm probability derived previously.

Fig. 1 illustrates the PDs of the GLRGDD and AMGDD detectors across varying SNRs. In the legend, ``$\circ$'' and ``$ \square $'' denote the theoretical results, while the dashed and dot-dashed lines depict the theoretical outcomes. For the simulation results, the $(k_1,k_2)$-th element of the noise covariance matrix is modeled as $\textbf{R}(k_1,k_2)=0.95^{|k_1-k_2|}$, $k_1,k_2=1,2,\cdots,O$. To alleviate the computational complexity, the PFA is set to be $10^{-3}$. To obtain the detection threshold, $10^5$ Monte Carlo simulations are run, and to obtain the PD, $10^4$ Monte Carlo simulations are run.
The results in Fig. 1 demonstrate that the theoretical results can be well matched with the simulation results, which confirms the precision of the statistical analysis results presented in this letter. Additionally, the detection probability (PD) of the GLRGDD exceeds that of the AMGDD.


By comparing the results shown in Fig. 1 and Fig. 2, it can also be noticed that with the increase in the number of row subspace dimension $P$, the PD of a detector also goes up. The reason is that when $P$ increases while the row subspace dimension $Q$ stays unchanged, the row information of the signal gets more concentrated, which improves the detection performance.

\begin{figure}[!htp]
\centering
\includegraphics[width=0.5\textwidth]{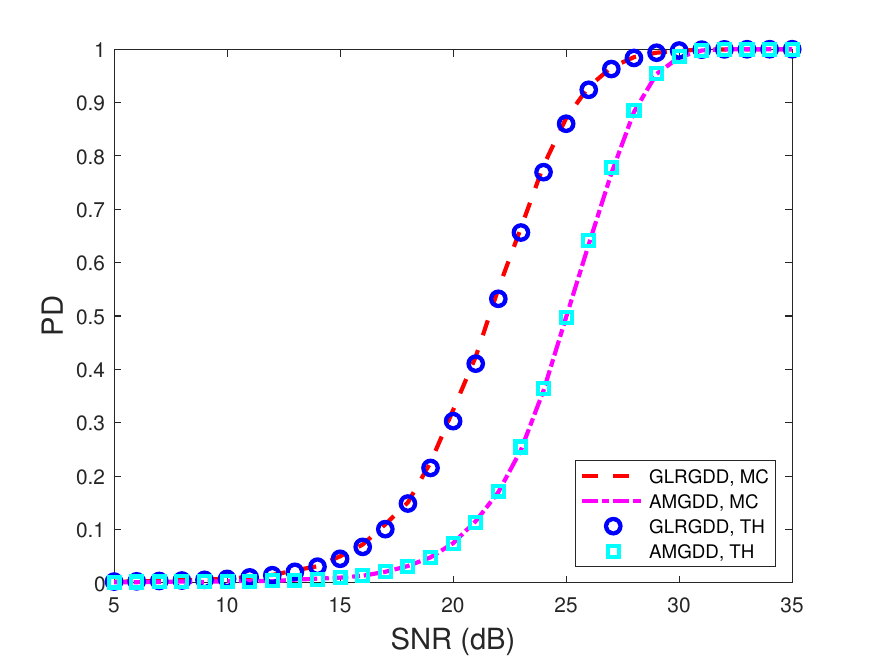}\\
\caption{PDs of the GLRGDD and AMGDD. $O=12$, $J=1$, $Q=3$, $L=11$, and $P=6$.}
\label{fig1}
\end{figure}

\begin{figure}[!htp]
  \centering
  \includegraphics[width=0.5\textwidth]{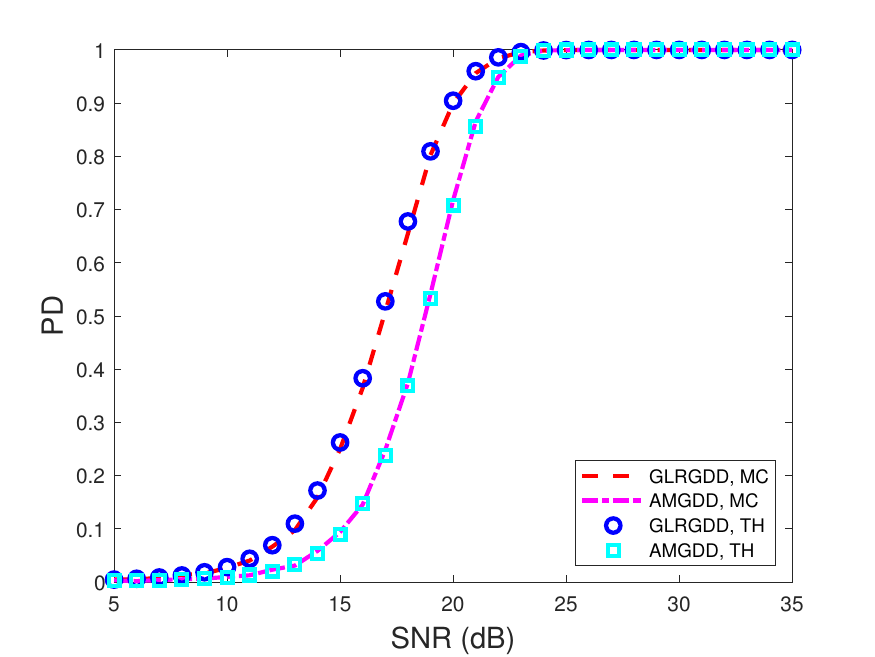}\\
\caption{PDs of the GLRGDD and AMGDD. $O=12$, $J=1$, $Q=3$, $L=11$, and $P=9$.}
  \label{fig:PDvSNR}
\end{figure}

Before closing this section, we would like to give some analysis about the computational complexities of the two detectors.
It can be seen from (5) that the main computational complexity of the detector GLRGDD comes from the construction and inverse operation of the matrix $\textbf{S}_+$. For the construction of $\textbf{S}_+$, its computational complexity is $\text{O}(O^2P + OP^2)$. For the inverse operation of $\textbf{S}_+$, its computational complexity is $\text{O}(O^3)$. Therefore, the computational complexity of the GLRGDD is $\max\left(\text{O}(O^2P + OP^2), \text{O}(O^3)\right)$.
Similarly, it can be seen from (7) that the main computational complexity of the detector AMGDD comes from the inverse operation of $\textbf{S}$. Therefore, the main computational complexity of the detector AMGDD is $\text{O}(O^3)$.
Based on the above analysis, it can be seen that when $P > O$, the computational complexity of the GLRGDD is greater than that of the AMGDD. In contrast, when $P \leq O$, the computational complexity of the detector GLRGDD is comparable to that of the detector AMGDD.

\section{Conclusions}

In this letter, we have deduced the statistical characteristics of the GLRGDD and AMGDD under the condition of a known spatial steering vector. The findings reveal that their statistical distributions encompass the complex F distribution and complex Beta distribution. Notably, the complex F distributions associated with both detectors share the same DOF, whereas the complex Beta distributions involved in each detector exhibit different DOFs. Based on these statistical distributions, we have obtained analytical expressions to calculate the PDs and PFAs for both detectors. Simulation results have confirmed the correctness of our theoretical derivations. Furthermore, the GLRGDD demonstrates superior detection performance to the AMGDD.

{\small
\bibliographystyle{IEEEtran}
\bibliography{D:/LaTexReference/Detection}
}
\end{document}